\documentclass[aps,prl,showpacs,twocolumn,amsmath,amssymb]{revtex4-1}
\usepackage{graphicx,bbm}

\newcommand{\be}{\begin{equation}}
\newcommand{\ee}{\end{equation}}

\newcommand{\ave}[1]{{\langle #1\rangle}}
\newcommand{\ii}{ {\rm i} }
\newcommand{\dd}{ {\rm d} }

\newcommand{\ZZ}{\mathbb{Z}}
\newcommand{\RR}{\mathbb{R}}
\newcommand{\CC}{\mathbb{C}}

\newcommand{\y}{{\rm y}}
\newcommand{\x}{{\rm x}}
\newcommand{\z}{{\rm z}}

\def\tr{{\,{\rm tr}}}

\def\one{\mathbbm{1}}

\newcommand{\AAA}{{\mathfrak A}}

\def\ave#1{\langle #1 \rangle}
\def\ii{{\rm i}}

\def\z{{\rm z}}

\def\tit#1{}

\begin{document}

\title{Lower bounds on high-temperature diffusion constants from quadratically extensive almost conserved operators}
 
\author{Toma\v z Prosen}
\affiliation{Department of Physics, Faculty of Mathematics and Physics, University of Ljubljana, Ljubljana, Slovenia}

\date{\today}

\begin{abstract}
We prove a general theorem which provides a strict lower bound on high-temperature Green-Kubo diffusion constants in locally interacting quantum lattice systems, under the assumption of existence of a quadratically extensive almost conserved quantity --- an operator whose commutator
with the lattice Hamiltonian is localized on the boundary sites only. We explicitly demonstrate and compute such a bound in two important models in one dimension, namely in the (isotropic) Heisenberg spin 1/2 chain and in the fermionic Hubbard chain.
\end{abstract}

  
\maketitle

{\em Introduction.--} Deriving or rigorously establishing diffusive transport in interacting extended systems from microscopic equations of motion is one of the main unsettled issues of statistical and theoretical condensed matter physics \cite{affleck}.
Even within the common and usually very realistic assumption of linear response, where the transport coefficients can be expressed via Green-Kubo  formulae \cite{green,kubo} in terms of integrated spatio-temporal current-current correlation functions at equilibrium \cite{mahan,evans}, the problem remains generally completely open as it is -- at least in the quantum domain -- connected to the infamous intractability of long-time many-body dynamics.

In mid 1990's a complementary version of the problem has been addressed in one-dimensional systems (locally interacting chains), namely conditions for establishing a ballistic transport, where the currents do not scale down with increasing the system size,
have been formulated by facilitating a Mazur bound \cite{mazur} on the so-called Drude weights \cite{czp} whose linear response expression can in turn be estimated in terms of local conserved quantities \cite{znp}.
Some delicate but important issues on correct treatment of boundary conditions and order of the two limits, the thermodynamic limit (TL) of size $n\to\infty$ and time $t\to\infty$, have only recently been resolved \cite{ip}.
The key feature to understand these issues is the emerging causality which arises in locally interacting quantum (or classical) gasses and is
efficiently encoded in terms of Lieb-Robinson bounds (LRB) \cite{lr} (which, although proposed more than forty years ago, have only recently
found truly remarkable and fundamental applications \cite{h04,h07}). This establishes a firm link between ballistic transport and integrability.

Furthermore, it has been realized \cite{znp} that local conserved quantities given by algebraic Bethe ansatz \cite{gm} are often irrelevant for the 
transporting quantity (e.g., are orthogonal to spin current in the zero magnetization sector, or charge current in the half-filled sector). On the other hand, studying the corresponding quantum master equation for the non-equilibrium integrable system driven by dissipative boundaries
reveals independent (almost) conserved quantities which are naturally related to a driven current and thus yield (nearly) optimal ballistic bounds  \cite{p11,pi}.

In this Letter we derive an inequality providing a strict lower bound of the Green-Kubo expression of the high-temperature diffusion constant in terms of a Lieb-Robinson group velocity of the lattice gas and a
quadratically extensive conserved operator $Q$, whose high-temperature square-expectation value scales as $\ave{Q^2} \propto n^2$. It is shown that such operators  naturally emerge from solving certain integrable boundary driven Lindblad equations by working out two important examples, namely the isotropic Heisenberg spin 1/2 chain and the fermionic Hubbard chain. We thus show rigorously that the spin transport in Heisenberg chains and spin and charge transport in fermionic Hubbard chains at high-temperature cannot be sub-diffusive (e.g. insulating).

{\em Inequality for diffusion constants.--} 
To define the necessary prerequisites, we follow notation of Ref.~\cite{ip} (also \cite{br}) and take $N \in \ZZ_+$ to be a local Hilbert space dimension, say $N=2$ for spins 1/2 or qubits,
and $\AAA_x \equiv \CC^{N \times N}$, $x \in \ZZ$, a local on-site matrix algebra.
We associate a matrix algebra $\AAA_{[x,y]}$, equipped with operator norm $\|\bullet\|$, to any finite open lattice of integers $[x,y] = \{x,x+1,\ldots,y-1,y\}$.
For convenience we consider our system to be defined on a one-dimensional lattice $\Lambda_n = [-n/2,n/2] \subset\ZZ$ of $|\Lambda_n|=n+1$ sites ($n$ assuming to be even for simplicity of notation), whereas generalization to $d-$dimensional lattices will be straightforward. Let the {\em local interaction} $h\in \AAA_{[0,d_{h}-1]}$ be an element of spin algebra on $d_h$ sites defining the translationally invariant Hamiltonian over the lattice $\Lambda_n$
\begin{equation}
H_{\Lambda_{n}}=\sum_{x=-n/2}^{n/2-d_{h}+1}h_{x},
\end{equation}
being a sum of local energy densities 
$h_{x}=\eta_{x}(h) \in \AAA_{[x,x+d_h-1]}$ obtained by a group of lattice shift morphisms  defined by $\eta_{y}(a_{x})=a_{x+y}$.
We shall always consider the high-temperature limit of observables first, thus we define an infninite-temperature Gibbs (tracial) state over 
$\AAA_{\Lambda_n}$ as $\ave{a} = (\!\tr\,a)/(\!\tr \one)$ satisfying $\ave{ab}=\ave{ba}$.
Furthermore, we assume existence of an {\em almost conserved} operator $Q_{\Lambda_n} \in \AAA_{\Lambda_n}$ satisfying
\begin{equation}
[H_{\Lambda_n},Q_{\Lambda_n}] = B_{\partial\Lambda_n} := b_{n/2} - b_{-n/2},
\label{almost}
\end{equation}
where $b_x \in \AAA_x$ is a local operator, which is {\em quadratically extensive}, namely the following limit exists and is finite
\begin{equation}
q:= \lim_{n\to\infty} \frac{\ave{Q_{\Lambda_n}^2}}{n^2}  \in (0,\infty).
\label{q}
\end{equation}
Finally, writing dynamics over $\AAA_{\Lambda_n}$ as $\tau_t(a) = e^{\ii H_{\Lambda_n}t} a e^{-\ii H_{\Lambda_n}t}$
we state the fundamental LRB \cite{lr,br,nos} as formulated in Ref.~\cite{bhv}:
Consider a sublattice $\Gamma \subset \Lambda_n$ with the complement $\Gamma^{\rm c} = \Lambda_n\setminus \Gamma$
defining a projection map over $\AAA_{\Lambda_n}$ as $(a)_\Gamma := \frac{ \tr_{\Gamma^{\rm c}}(a)}{ \tr_{\Gamma^{\rm c}}(\one)} \otimes \one_{\Gamma^{\rm c}}$, 
and take a local observable $f\in\AAA_X$ over a subset $X\subset \Gamma$ of $|X|$ sites. LRB then states that the time evolution projected to $\Gamma$ approximates the exact evolution exponentially well within a cone which opens at $X$ at some speed.
Namely,  for {\em any} $\mu > 0$, there exist constants $C_\mu > 0,v_\mu > 0$, independent of $f,X,\Gamma,t$ and $n$, such that
\begin{equation}
\| \tau_t(f) - (\tau_t(f))_{\Gamma}\| \le C_\mu |X| \| f\| e^{-\mu ({\rm dist}(X,\Gamma^{\rm c}) - v_\mu |t|)},
\label{eq:BHV}
\end{equation}
where ${\rm dist}(X,Y) = \min_{x\in X,y\in Y}|x-y|$. The optimized bound defines the group velocity as $v=\inf_{\mu>0}v_\mu$ which in turn
can be estimated from above in terms of the strength of interactions, for example
for one-dimensional lattice and nearest-neighbor interactions, $d_h=2$, one can estimate \cite{bhv,them}, 
\begin{equation} 
v \le 6 \| h' \|,
\label{vest}
\end{equation} 
where $h'$ is the pure interaction part of $h$, with 1-site terms (external fields, or on-site interactions) subtracted.

We are now in position to state the main result: \\
{\bf Theorem:} {\em Take an arbitrary self-adjoint element of local algebra $j \in \AAA_{[0,d_j-1]}$ satisfying $\ave{j}=0$,
and define a spatiotemporal correlation function of the infinite lattice dynamics as 
\begin{equation}
C(x,t) = \lim_{n\to\infty} \ave{ \eta_x(j) \tau_t(j)}.
\end{equation}
Assuming that $C(t) := \sum_{x=-\infty}^\infty C(x,t)$ exists for any $t$, that $D:=\int_{-\infty}^\infty \!\dd t\,C(t)$ and $D':=\int_{-\infty}^\infty\!\dd t \,|t|C(t)$ exist as well, and that $Q_{\Lambda_n}$ has a well defined component along $j$, $Q^j := \lim_{n\to\infty} \ave{j Q_{\Lambda_n}}$, 
 the following inequality holds}
\begin{equation}
D \ge \frac{|Q^j|^2}{8 v q}.
\label{ineq}
\end{equation}
Note that $D$ is just a Green-Kubo diffusion constant of the conserved quantity associated with the current density $j$, e.g., magnetization, charge, or energy (heat) \cite{mahan,evans} and is related to the corresponding high-temperature d.c. conductivity $\sigma$ via the Einstein relation $\sigma = D/T$.

\begin{figure}
\centering	
\vspace{-1mm}
\includegraphics[width=\columnwidth]{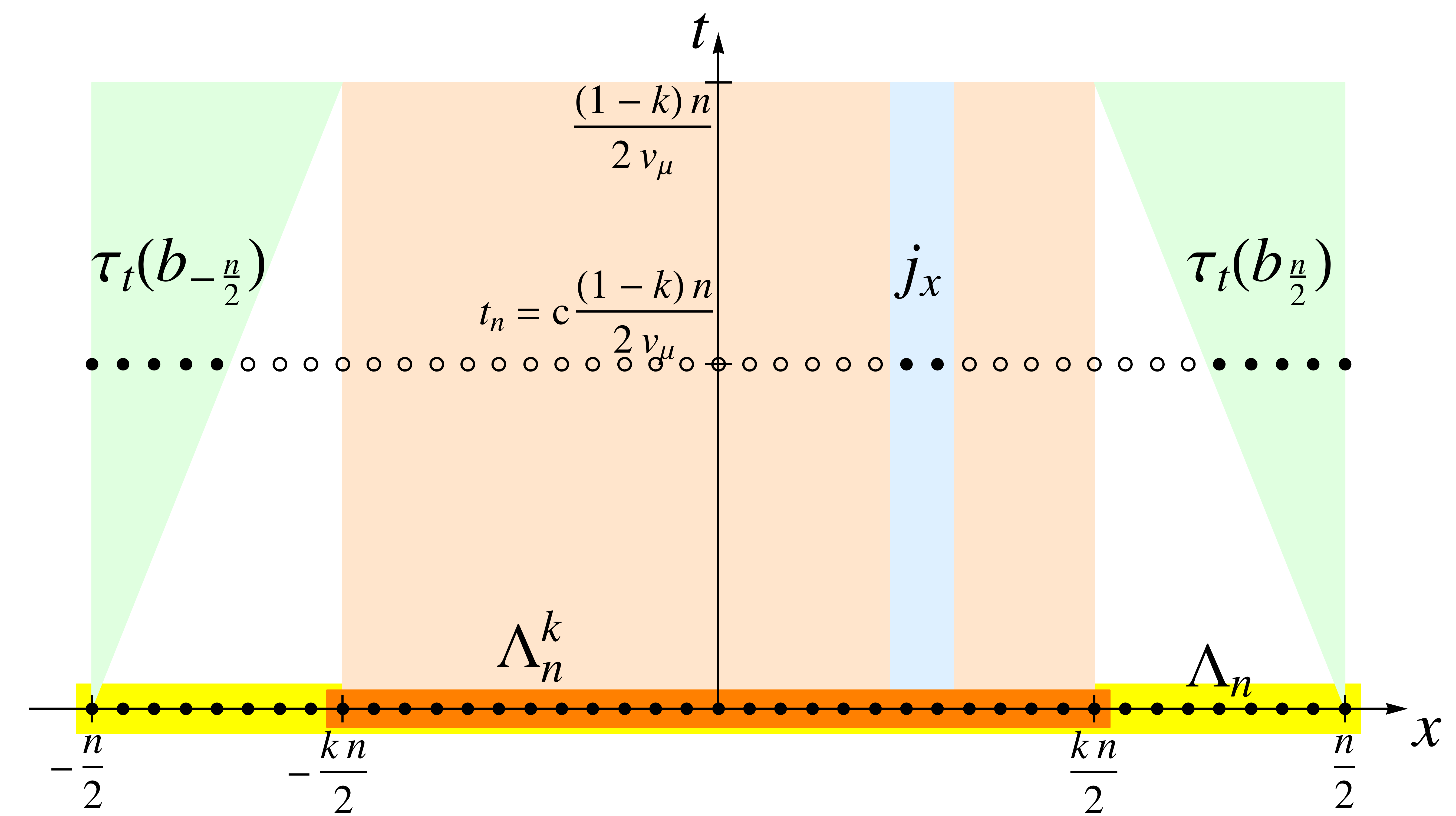}
\vspace{-1mm}
\caption{(Color online) Illustrating locality of spatio-temporal dynamics for estimation of the second term of (\ref{AA}).}
\label{fig}
\end{figure}

\noindent
{\bf Proof:} To establish (\ref{ineq}) we slightly modify the proof for bounding the Drude weights \cite{ip}.
Let us consider a scaling factor $k\in (0,1)$, define a scaled-down lattice $\Lambda^k_n = [-\lfloor kn/2\rfloor,\lfloor kn/2\rfloor]$, and
write an extensive current as $J_{\Lambda^k_n} =\sum_{x=-\lfloor kn/2\rfloor }^{\lfloor kn/2\rfloor-d_j+1} j_x$. Then we
study the following finite-time-averaged self-adjoint operator
\begin{equation}
A_{n,t}:=\frac{1}{t}\int_{0}^{t}\dd t' \left(\tau_{t'}(J_{\Lambda^k_{n}})-\frac{\alpha}{n} Q_{\Lambda_{n}}\right),
\label{eqn:ansatz}
\end{equation}
where  $\alpha \in \mathbb{R}$ is a free parameter. Since $A^2_{n,t} \ge 0$, we have $\ave{A^2_{n,t}}\ge 0$ for any $t,\alpha\in\RR$, $n\in\ZZ^+$, i.e.
\begin{eqnarray}
&& \int_{0}^{t}\dd t'\int_{0}^{t}\dd t'' \frac{1}{t^2}\ave{\tau_{t'}(J_{\Lambda^k_{n}})\tau_{t''}(J_{\Lambda^k_{n}})} -\nonumber \\
&& -  \int_0^t \dd t' \frac{2\alpha}{n t} \ave{\tau_{t'}(J_{\Lambda^k_{n}})Q_{\Lambda_{n}}} +\frac{\alpha^{2}}{n^2}\ave{Q^{2}_{\Lambda_{n}}} \ge 0. \label{AA}
\end{eqnarray}
Fixing exponent $\mu>0$, we introduce another scaling factor $c\in(0,1)$ and write the size dependent time-span as $t=t_n := \frac{c(1-k) n}{2 v_\mu}$ (see Fig.~\ref{fig}). We proceed to show that the TL $n\to\infty$ of (\ref{AA}), at fixed $k,c$, exists term by term, evaluating them below.
Existence of infinite lattice dynamics together with LRB guarantees homogeneity of correlation functions in the entire scaled domain 
(c.f. Theorem 6.2.11 of Ref.~\cite{br}), namely
$$|C(x'-x'',t'-t'') - \ave{\tau_{t'}(j_{x'}) \tau_{t''}(j_{x''})}| \to 0$$ for $n\to\infty$ uniformly in space-time domain $|x'|,|x''| < \frac{kn}{2}$ and $|t'-t''| < t_n$, hence the {\em first} term of (\ref{AA}) evaluates to
$$\frac{2 k v_\mu}{c (1-k)}\int_{-t_n}^{t_n} \dd t' \left(1-\frac{|t'|}{t_n}\right)C(t') \to \frac{2 k v_\mu}{c(1-k)} D$$ due to existence of $D'$.
The {\em last} term of (\ref{AA}), via (\ref{q}), in TL evaluates to $\to \alpha^2 q$. The integrand of the {\em second} term of (\ref{AA}) 
$ \ave{\tau_{t'}(J_{\Lambda^k_{n}})Q_{\Lambda_{n}}} =  \ave{J_{\Lambda^k_{n}}\tau_{-t'}(Q_{\Lambda_{n}})}$ can be first estimated 
by solving equation of motion  (\ref{almost}) for
$\dd Q_{\Lambda_n}/\dd t = \ii b_{n/2} - \ii b_{-n/2}$, namely 
$$
\tau_{-t'}(Q_{\Lambda_n}) = Q_{\Lambda_n} - \ii \int_{-t'}^0 \dd t'' (\tau_{t''}(b_{n/2}) - \tau_{t''}(b_{-n/2})).
$$
This implies 
\begin{eqnarray*}
&&|\ave{\tau_{t'}(J_{\Lambda^k_{n}})Q_{\Lambda_{n}}} - \ave{J_{\Lambda^k_n}Q_{\Lambda_n}}| \le \\ 
&&\sum_{x=-\lfloor kn/2\rfloor}^{\lfloor kn/2\rfloor - d_j+1}\int_{-t'}^0\dd t''
(|\ave{j_x \tau_{t''}(b_{n/2})}|+|\ave{j_x \tau_{t''}(b_{-n/2})}|).
\end{eqnarray*}
Defining $\Gamma = \Lambda_n\setminus [x,x+d_j-1]$, each term on the RHS can be further estimated using LRB (\ref{eq:BHV}),
$|\ave{j_x \tau_{t''}(b_{-n/2})}| \le |\ave{j_x (\tau_{t''}(b_{-n/2}))_{\Gamma}}| + \| j\| \|\tau_{t''}(b_{-n/2})-(\tau_{t''}(b_{-n/2}))_{\Gamma}\|
\le C_\mu \| j \| \| b\| e^{-\mu (x - v_\mu t'')}$. As the infinite-temperature state is separable, we have $\ave{j_x (\tau_{t''}(b_{-n/2}))_{\Gamma}}=0$.
Estimating similarly the other terms with $b_{n/2}$ we arrive after summation over $x$ and integration over $t''$ at the estimate
\begin{eqnarray*}
|\ave{\tau_{t'}(J_{\Lambda^k_{n}})Q_{\Lambda_{n}}}\!-\!\ave{J_{\Lambda^k_n}Q_{\Lambda_n}}| \le \frac{4\| j\| \| b\| C_\mu e^{-\frac{\mu}{2}(1-k)(1-c)n}}{(1-e^{-\mu})\mu v_\mu}.
\end{eqnarray*}
This gives a uniform convergence of the integral in the {\em second} term of (\ref{AA}), which finally $\to 2\alpha k Q^j$ as $n\to\infty$.

Collecting TL of all three terms we get an inequality
\begin{equation}
\frac{2 k v_\mu}{c(1-k)} D - 2 \alpha k Q^j + \alpha^2 q \ge 0, 
\end{equation}
which, after optimizing with respect to $\alpha$, yields $$D \ge \frac{c k (1-k)}{2v_\mu} \frac{(Q^j)^2}{q}.$$ After taking a supremum of the RHS, 
$\sup_{\mu,k,c} \frac{c k (1-k)}{2v_\mu} = \frac{1}{8v}$, the inequality (\ref{ineq}) is recovered.

We note that our constructions should straightforwardly generalize to higher-dimensional lattices, where $B_{\partial \Lambda_n}$ (\ref{almost}) should now be understood
as an `area operator' localized anywhere on the boundary of the lattice. The only difference is that the prefactor $1/8$ in (\ref{ineq}) should then be replaced by another numerical prefactor
depending only on dimensionality and geometry of the lattice.

We shall now proceed to demonstrate two nontrivial implementations of the bound (\ref{ineq}).

{\em Diffusion bound in the Heisenberg chain.--} In the isotropic spin 1/2 Heisenberg chain, $N=2$, where the local spin algebra is generated by Pauli matrices $\sigma^{\x,\y,\z}$, 
$\sigma^\pm = \frac{1}{2}(\sigma^\x \pm \ii \sigma^\y)$, and Hamiltonian density reads 
$
h=J (2\sigma^+\otimes \sigma^- + 2 \sigma^-\otimes\sigma^+ + \sigma^\z \otimes\sigma^\z + \one),
$
with $d_h=2$, all attempts of establishing clearly whether the high-temperature spin transport with the current density $j = 4\ii J (\sigma^+ \otimes \sigma^- - \sigma^- \otimes \sigma^+)$, satisfying continuity equation $\ii [H_{\Lambda_n},\sigma^\z_x] = j_{x}-j_{x-1}$, is diffusive, ballistic, or insulating, failed \cite{peter1,fabian,peter2,affleck}. The most accurate numerical evidence comes from time-dependent density matrix renormalization group simulations in Liouville space \cite{z11} suggesting that the transport is anomalous with the steady-state current $\ave{j} \sim n^{-1/2}$ which would imply super-diffusion $D=\infty$.
Analytic study of non-equilibrium steady state density operator for the boundary driven Linbdlad equation over the lattice $\Lambda_n$, with jump (Lindblad) operators $\sigma^+_{-n/2}$ and $\sigma^-_{n/2}$
resulted \cite{p11} in the following almost conserved operator
\begin{equation}
Z_{\Lambda_n} = \sum_{x,y\in\Lambda_n}^{x < y} \sigma^+_x \sigma^-_y
\end{equation}
satisfying $[H_{\Lambda_n},Z_{\Lambda_n}] = -J \sigma^\z_{-n/2} + J \sigma^\z_{n/2}$.
Writing a self-adjoint combination $Q_{\Lambda_n} = \ii(Z_{\Lambda_n}-Z^\dagger_{\Lambda_n})$ we calculate straightforwardly 
$\ave{Q_{\Lambda_n}^2} = \frac{1}{4}n(n-1)$, yielding $q=1/4$, and $Q^j = 2J$. Together with $\| h\|= 2|J|$, inequalities (\ref{ineq}) and (\ref{vest}) result in a rigorous lower bound on spin-diffusion constant: 
$D \ge |J|/6 $. This bound is likely highly non-optimal in view of the behavior suggested numerically in Ref.~\cite{z11}, and is safely consistent with older arguments \cite{muller,mccoy}, but it is the first rigorous result of this sort.
However, the next example perhaps gives a more useful bound, tighter to best empirical evidence.

{\em Diffusion bound in the Hubbard chain.--} Consider a fermionic single band Hubbard chain \cite{book} over $\Lambda_n$ which after introducing the Jordan-Wigner transformation 
$c_{\uparrow,x}=P^{(\sigma)}_{-n/2,x-1} \sigma_x^-$ and $c_{\downarrow,x}=P^{(\sigma)}_{-n/2,n/2} P^{(\tau)}_{-n/2,x-1} \tau_x^-$, where $P^{(\sigma)}_{x,y}:=\sigma_{x}^{\rm z} \sigma^\z_{x+1} \cdots \sigma_y^{\rm z}$, 
$P^{(\tau)}_{x,y}:=\tau_{x}^{\rm z}\tau_{x+1}^{\rm z}\cdots \tau_y^{\rm z}$, 
maps exactly to a spin 1/2 ladder described by two sets of Pauli matrices at each site, $\sigma^{\pm,\z},\tau^{\pm,\z}$ spanning local dimension $N=4$. The Hamiltonian density for hopping $t$ and interaction $U$ reads
$h = h' + \frac{U}{8}(\sigma^\z \tau^\z\otimes \one + \one\otimes\sigma^\z \tau^\z)$, with $h'= t (\sigma^+\otimes \sigma^- + \sigma^-\otimes \sigma^+ + \tau^+\otimes \tau^- + \tau^-\otimes\tau^+)$, $d_h=2$.
Here we consider charge and spin currents $j^{\rm c,s} = -2\ii t \left[\sigma^+\otimes\sigma^- - \sigma^-\otimes\sigma^+ \pm (
\tau^+\otimes\tau^- - \tau^-\otimes\tau^+)\right]$, satifying continuity equations $\ii [H_{\Lambda_n},\sigma^\z_x \pm \tau^\z_x] = j^{\rm c,s}_x - j^{\rm c,s}_{x-1}$.
It has recently been shown \cite{p13} that boundary driven steady-state  Lindblad equation for the Hubbard chain with charge-driving 
Lindblad operators $\sigma^+_{-n/2},\tau^+_{-n/2},\sigma^-_{n/2},\tau^-_{n/2}$ is exactly solvable, which in the leading order in the system-bath coupling yields a quadratically extensive operator
\begin{eqnarray}
Z^{\rm c}_{\Lambda_n} &=& -\frac{t}{2}\sum_{x=-n/2}^{n/2-1}(\sigma^+_x\sigma^-_{x+1}+\tau^+_x\tau^-_{x+1})  \\
&+&  \frac{U}{2}\sum_{x,y\in{\Lambda}_n}^{x<y}(-1)^{x-y} \sigma^+_x P^{(\sigma)}_{x+1,y-1} \sigma^-_y \tau^+_x P^{(\tau)}_{x+1,y-1} \tau^-_y, \nonumber
\end{eqnarray}
where $P^{(\sigma,\tau)}_{x,y}\equiv\one$ if $x > y$, satisfying the almost conservation condition
$[H_{\Lambda_n},Z_{\Lambda_n}] = \frac{t^2}{4}(\sigma^\z_{-n/2} + \tau^\z_{-n/2} - \sigma^\z_{n/2} - \tau^\z_{n/2}$).
Similarly, the leading order fixed point of spin-driven Lindblad dynamics with jump operators
 $\sigma^+_{-n/2},\tau^-_{-n/2},\sigma^-_{n/2},\tau^+_{n/2}$, yields another almost conserved quadratically extensive operator
 \begin{eqnarray}
 Z^{\rm s}_{\Lambda_n} &=& -\frac{t}{2}\sum_{x=-n/2}^{n/2-1}(\sigma^+_x\sigma^-_{x+1}+\tau^-_x\tau^+_{x+1})  \\
&+&  \frac{U}{2}\sum_{x,y\in{\Lambda}_n}^{x<y} \sigma^+_x P^{(\sigma)}_{x+1,y-1} \sigma^-_y \tau^-_x P^{(\tau)}_{x+1,y-1} \tau^+_y, \nonumber
\end{eqnarray}
satisfying $[H_{\Lambda_n},Z^{\rm s}_{\Lambda_n}] = \frac{t^2}{4}(\sigma^\z_{-n/2} - \tau^\z_{-n/2} - \sigma^\z_{n/2} + \tau^\z_{n/2})$.
Writing the following current-like self-adjoint combinations $Q^{\rm c,s}_{\Lambda_n} = \ii (Z^{\rm c,s}_{\Lambda_n}-(Z^{\rm c,s}_{\Lambda_n})^\dagger)$,
we calculate straightforwardly: $\ave{(Q^{\rm c,s}_{\Lambda_n})^2} = n \frac{t^2}{4} + n (n-1)\frac{U^2}{64}$, hence $q = U^2/64$,
and $Q^j = \ave{j^{\rm c,s} Q^{\rm c,s}_{\Lambda_n}} = t^2$. Finally, considering $\| h'\| = 2|t|$ we obtain a strict, uniform bound on high-temperature spin and charge diffusion constant 
\begin{equation}
D^{\rm c,s} \ge \frac{ 2|t|^3}{3 U^2}.
\end{equation}
This bound is qualitatively and quantitatively consistent with the suggested diffusive spin and charge transport based on numerical simulations of the Liouville-space Lindblad dynamics in the linear-response regime \cite{pz12}.

{\em Discussion.--} Our bounds on Green-Kubo diffusion 
coefficients will rarely come close to tight due to a generally large overestimation of the (Lieb-Robinson) group velocity in interacting lattice gases.
Nevertheless, these bounds disclose a fundamental importance of the newly emerging quadratically extensive almost conserved quantities --- in the absence of linearly extensive (local) relevant conserved quantities --- for understanding the possibility of a diffusive transport in exactly solvable lattice gases. Here we have elaborated on two fundamental quantum chains, Heisenberg and Hubbard.  We note that in these two cases the quadratically extensive conserved quantities can also be interpreted as the (truncated, finite-size) generators of the {\em Yangian symmetry} of the model \cite{bernard,korepin}. We believe our results should be straightforwardly extended to classical integrable lattice gases as well where a possibility of diffusive transport has recently been suggested \cite{pz13}.  So far, our theorem applies only to a high-temperature limit, whereas extension  to arbitrary temperatures is expected to be feasible in one dimension where exponential clustering (decay of correlation) of the Gibbs state should should step in instead of separability of the tracial (infinte temperature) state.

Useful remarks by Enej Ilievski and support by the grants P1-0044 and J1-5439 of Slovenian Research Agency (ARRS) are gratefully acknowledged.

\end{document}